\documentclass[pra,aps,showpacs,twocolumn]{revtex4}
\begin{document}
\title{Optical implementations,  oracle equivalence, and the
Bernstein-Vazirani algorithm}         
\author{Arvind}       
\email{arvind@quantumphys.org}
\affiliation{Department of Physics, 
Indian Institute of Technology Madras,
Chennai 600036}
\altaffiliation[Also at:]{ 
Department of Physics, Guru Nanak Dev 
University, Amritsar 143005}
\author{Gurpreet Kaur}
\affiliation{Department of Physics, 
IIT Madras,
Chennai 600036}
\author{Geetu Narang}
\affiliation{Department of Physics, Guru Nanak Dev University,
Amritsar, 143 005}
\pacs{03.67.Lx,42.25.Ja,42.25.Hz}
\begin {abstract}

We describe a new implementation of the Bernstein-Vazirani
algorithm which relies on the fact that the polarization states
of classical light beams can be cloned.  We explore the
possibility of computing with  waves and discuss a classical
optical model capable of implementing any algorithm (on $n$
qubits) that does not involve entanglement.  The
Bernstein-Vazirani algorithm (with a suitably modified oracle),
wherein a hidden $n$ bit vector is discovered by one oracle query
as against $n$ oracle queries required classically, belongs to
this category.  In our scheme, the modified oracle is also
capable of computing $f(x)$ for a given $x$, which is not
possible with earlier versions used in recent NMR and optics
implementations of the algorithm. 

\end{abstract}
\maketitle 
\section{Introduction}
\label{introduction}

Quantum mechanical systems have a large in-built information
processing ability and can hence be used to perform
computations~\cite{divin-sc-95,benn-pt-95,benn-nat}.
The basic unit of quantum information is the \underline{qu}antum
\underline{bit} (qubit), which can be visualized as a quantum
two-level system.  The implementation of quantum logic gates is
based on reversible logic and the fact that the two states of a
qubit can be mapped onto logical 0 and
1~\cite{bar-pr-95,divin-pr-95,divin-roy-98}.  The
quantum mechanical realization of logical operations can be used
to achieve a computing power far beyond that of any classical
computer~\cite{feyn-the-82,benioff,deu-roy-85,deu-roy-89}.
A few quantum algorithms have been designed and experimentally
implemented, that perform certain computational tasks
exponentially faster than their classical counterparts. While the
Deutsch-Jozsa (DJ) algorithm ~\cite{deu-roy-92} and Shor's
quantum factoring algorithm~\cite{shor-siam-97,ekert-revmod-96}
lead to an exponential speedup, Grover's rapid search
algorithm~\cite{grover-prl-97} and the
Bernstein-Vazirani~\cite{BV-97} algorithm are examples where a
substantial (though non-exponential) computational advantage is
achieved.

The exponential gain in computational speed achieved by quantum
algorithms is intimately related to
entanglement~\cite{MHorodecki01}, and it turns out that when
there is no entanglement (or the amount of entanglement is
limited) in a pure state, the dynamics of a quantum algorithm can be
simulated efficiently via classically deterministic or classical
random means~\cite{jozsabook,nielsen,JL02}.  However in
algorithms that do not lead to an exponential gain in speed or in
those that use mixed states, the possibilities of achieving
speedup without entanglement still exist~\cite{tal-mor-qph}.

Classical waves share certain properties of quantum systems.  For
example, the polarization states of a beam of light can act as
qubits.  It is to be noted that the superposition of classical
waves does not lead to entanglement. For $n$ beams of light, with
their polarization states providing us with $n$ qubits, we can
only implement $U(2)\otimes U(2) \cdots \otimes U(2)$ transformations
via optical elements~\cite{simon-pla-90} and cannot in general
implement $U(2^n)$ transformations. Therefore, although
superposition and interference are present and can be utilized,
their scope is limited compared to what could be achieved with
$n$ qubits which are actually quantum in character. However, it
is interesting to explore the question if any useful computation
could be performed  with classical waves, that exploits their
superposition and interference. It turns out that, if an
algorithm based on qubits does not involve entanglement at any
stage of its implementation, it can be realized using this
classical model.  The Deutsch-Jozsa algorithm for one and two
qubits and the Bernstein-Vazirani algorithm for any number of
qubits, can be re-cast in this form with a suitable modification
of the oracle~\cite{Meyer00soph,BV-97,TerhalSmolin98}.
This modification of the algorithm has been central to the
implementation of the Deutsch-Jozsa algorithm up to two qubits
~\cite{collins-pra-98,arvind-pramana,kavita-pramana} and the Bernstein-Vazirani
algorithm on any number of qubits using NMR~\cite{djnmr} as well
as optics~\cite{djoptics} and superconducting
nanocircuits~\cite{bvsuper}.

In this paper, we propose a model based on classical light
beams in which the $n$-qubit eigen states are mapped on to
polarization states of these beams, and un-entangling unitary
operators are implemented using passive optics. An added
feature in this model is that cloning of states is possible
as we are working entirely within the domain of classical
optics.  It turns out that this possibility of cloning along
with interference leads to interesting results for certain
algorithms.

We discuss an entirely new scheme for implementing the
Bernstein-Vazirani algorithm. Instead of Hadamard
transformations, we use cloning and re-interference to discover a
hidden $n$-bit binary vector $a$, using only a single oracle
call. The non-entangling nature of the modified oracle for the
Bernstein-Vazirani algorithm is central to this implementation as
it is for the earlier implementations~\cite{djnmr,djoptics}.
However this scheme differs from the earlier schemes in two ways:
(a) instead of Hadamard transformation we use the cloning of
classical beams via beam splitters, and (b) we are able to
operate the modified oracle in the `classical' mode as well,
wherein we are able to obtain $f(x)$ for a given $x$.

The material in this paper is arranged as follows: the optical
model based on polarization states is described in
Section~\ref{optics}. Section~\ref{BV-algo} begins with the
description of the original Bernstein-Vazirani algorithm and
later discusses the modified oracle and its implications.  The
new optical scheme is described in Section~\ref{new-algo} and
Section~\ref{conclusion} has some concluding remarks.
\section{Optical implementation based on polarization}
\label{optics}
Consider a classical system consisting of a monochromatic
light beam propagating in a given direction with a pure
polarization. The polarization states of such a beam are in
one-to-one correspondence with the states of a two-level
quantum system and the beam can therefore be visualized as a qubit.
The unitary transformations that transform one polarization state
to another can be easily performed.
Consider a
birefringent plate with its thickness adjusted to introduce
a phase difference of $\eta\/$ between the $x\/$ and $y\/$
components of the electric field, with its slow axis making
an angle $\phi\/$ with the $x\/$ axis.  The unitary operator
corresponding to this plate is given by
\begin{equation}
U(\eta,\phi) \!=\!
\left[\begin{array}{cc} \cos\phi & -\sin \phi\\
                         \sin\phi & \cos \phi\end{array}
\right]\!\!
\left[\begin{array}{lr} e^{i\eta/2} & \!\!\!\!\!\!0\\ 0 &
\!\!\!\!\!\!e^{-i \eta/2}
\end{array} \right]\!\!
\left[\begin{array}{cc} \cos\phi & \sin \phi\\
                         -\sin\phi & \cos \phi\end{array} \right]
\label{biref}
\end{equation}
For $\eta=\pi\/$, it becomes a half-wave plate  
(denoted by $H_{\phi}\/$), while for $\eta=\pi/2\/$ it becomes
a quarter-wave plate (denoted by $Q_{\phi}\/$).
It has been  shown that all $U(2)\/$ transformations can be 
realized on the polarization states by taking two quarter-wave
plates and one half-wave plate with suitable choices of angles of
their slow axes with the $x\/$ axis.  We will henceforth refer 
to this device, capable of implementing $SU(2)\/$
transformations, as ``Q-H-Q'' (a detailed discussion is found
in~\cite{simon-pla-90}). Combining this with an overall trivial 
phase transformation, we can implement the complete set of $U(2)$
transformations.
\unitlength=0.6mm
\thicklines
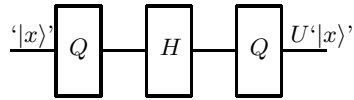
\begin{figure}[h]
\begin{picture}(100,50)(0,10)
\multiput(10,25)(20,0){4}{\line(1,0){10}}
\multiput(20,15)(20,0){3}{\line(0,1){20}}
\multiput(30,15)(20,0){3}{\line(0,1){20}}
\multiput(20,15)(20,0){3}{\line(1,0){10}}
\multiput(20,35)(20,0){3}{\line(1,0){10}}
\put(23,24){$Q$}
\put(43,24){$H$}
\put(63,24){$Q$}
\put(10,27){`$\vert x\rangle$'}
\put(72,27){$U$`$\vert x\rangle$'}
\end{picture}
\label{QHQ}
\caption{The action of the Q-H-Q device on the polarization state `$\vert
x\rangle$' of a single beam, taking it to a state `$\vert
x\rangle$', where $U$ could be an arbitrary $SU(2)$
transformation} 
\end{figure}

Further, let us map the $x\/$ polarization state to logical
$1\/$ and the $y\/$ polarization state to logical $0\/$.
With this mapping, we proceed to work with this system as a
qubit.  Since this system comprises essentially of classical
elements, we call it a ``classical qubit''. We will use a
notation where we specify the polarization state as `$\vert
x\rangle$' (i.e a ket vector within quotation marks)
throughout this paper, where $x$ can take values $0$ or $1$.
Multiple beams of this type can be considered and on each
one of them arbitrary $U(2)$ transformations can be
performed.  All the computational basis states  are mapped
to appropriate polarization states using the above mapping.
It is to be noted that we cannot obtain any entangled states
here because the transformations available are
$U(2)\otimes U(2)\cdots \otimes U(2)$.

\begin{figure}[h]
\thicklines
\unitlength=1mm
\begin{picture}(50,50)(10,10)
\put(10,30){\line(1,0){40}}
\put(30,10){\line(0,1){40}}
\put(25,25){\line(1,1){10}}
\put(12,32){`$\vert x\rangle$'}
\put(18,15){$e^{i\theta}$`$\vert x\rangle$'}
\put(38,32){$\sqrt{t}`\vert x\rangle'-e^{i\theta}\sqrt{r}`\vert x\rangle'$}
\put(32,50){$e^{i \theta}\sqrt{t}`\vert x\rangle'+\sqrt{r}`\vert x\rangle'$}
\put(31,37){\small BS}
\end{picture}
\label{BS-action}
\caption{The action of a beam splitter with transmission
coefficient $t$ and reflection coefficient $r$ on a classical light
beam with polarization state given by `$\vert x\rangle$'.
The same polarization is being sent into both ports of
the beam splitter and no polarization change occurs during the
whole process. For instance, if the beam in one of the ports is missing
and we use a 50-50 beam splitter, the beam splitter
generates two identical beams which are clones of the
original input beam and with their intensity reduced to half.}
\end{figure}
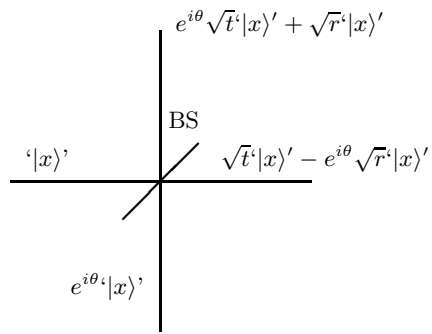
A beam splitter can be used to `split' a beam and also to
interfere beams with the same polarization. The transformation 
matrix of this operation on the amplitudes is given by 
\begin{equation}
\mbox{BS} = \left(\begin{array}{cc}
\sqrt{t} & -\sqrt{r}\\
\sqrt{r} & \sqrt{t}
\end{array}\right)
\end{equation}
Where $t$ and $r$ are transmission and reflection coefficients
respectively.
This matrix acts on the amplitudes of the two beams
entering the two ports of the beam splitter and not on the
polarization states. Polarization states do not undergo
any transformation 
under the action of the beam splitter.
\section{The Bernstein-Vazirani algorithm and a modified oracle}
\label{BV-algo}
Consider the binary
function $f(x)$ defined from an $n$-bit domain space to a $1$-bit range.  
\begin{equation}
f:\{0,1\}^n \longrightarrow \{0,1\}
\end{equation}
The
function is considered to be of the form $f(x)=a \cdot x$,
where $a$ is an $n$
bit string of zeros and ones and $a.x$ denotes bitwise
XOR( or scalar product modulo 2):
\begin{equation}
f(x) =a_{1}x_{1} \oplus a_{2}x_{2} \oplus ........a_{n}x_{n}
\label{bv_function}
\end{equation}

The aim of the algorithm is to find the $n$-bit string $a$,
given that we have access to an oracle which gives us the values
of the function $f(x)$ when we supply it with an input $x$.
Classically at least $n$ queries to the oracle are  required in order
to find the binary string $a$. The Bernstein-Vazirani algorithm
solves this problem with a single query to a quantum oracle of
the form 
\begin{equation}
\vert x \rangle_{n-\rm qubit} \vert y \rangle_{1-\rm qubit}
\buildrel{U_a}\over{\longrightarrow} \vert x \rangle_{n-\rm
qubit} \vert f(x)
\oplus y \rangle_{1-\rm qubit}
\label{oracle1}
\end{equation}
where $ x \in \{0,.........2^{n-1}\}$ is a 
data register and $ \vert y \rangle$ acts as 
a target register.
The algorithm works as follows: begin with an initial state
with the first $n$-qubits in $\vert 0 \rangle$ state and the last
qubit in the state $\vert 1 \rangle$. Apply a Hadamard
transformation on all the $n+1$ qubits and then make a call to the
oracle giving the following results:
\begin{eqnarray}
&&\vert 0 \rangle^n 
\vert 1 \rangle \buildrel{H^{\otimes
n+1}}\over{\longrightarrow}\frac{1}{2^{n/2}}
\sum_{x=0}^{2^{n}-1} \vert x \rangle
\;{1 \over \sqrt{2}} (\vert 0 \rangle - \vert 1 \rangle)\nonumber \\
&&\buildrel{U_a}\over{\longrightarrow} \frac{1}{2^{n/2}}
\sum_{x=0}^{2^{n}-1}  (-1)^{(x.a)} \vert x
\rangle \frac{1}{\sqrt{2}}
(\vert 0 \rangle - \vert 1 \rangle )\nonumber \\
&&\buildrel{H^{\otimes n+1}}\over{\longrightarrow}{1\over
2^n}\sum_{x=0}^{2^{n}-1}  
\sum_{z=0}^{2^{n}-1} (-1)^{(x.a)}(-1)^{(x.z)} \vert z
\rangle \vert 1 \rangle
\nonumber \\
&&\equiv \vert a \rangle
\vert 1 \rangle 
\label{bv-orig}
\end{eqnarray}
where we have used the fact that 
$$ {1\over2^n}\sum_{x=0}^{2^{n}-1} 
(-1)^{(a.x)}(-1)^{(x.z)} =
\delta_{a z} $$

A measurement in the computational basis immediately reveals
the binary vector $a$. This algorithm therefore achieves the
discovery of the vector $a$ in a single oracle call as
opposed to $n$ oracle calls required classically. The
oracle~(\ref{oracle1}) has been queried on a superposition
of states for this algorithm.  However, if we query the
oracle on a state $\vert x \rangle$ with the function
register set to $\vert 0 \rangle$ we will recover the value
$f(x)$ in the function register. This demonstrates that we
can run the oracle in the classical mode when desired.

\subsection{Oracle modification and implementation without
entanglement} This unitary oracle~(\ref{oracle1}) requires
$n+1$ qubits and can be operated in two different ways.  If
we use eigen states in the input and set $y=0$, the
algorithm outputs $f(x)$ for a given input $x$ in a
reversible manner (which the original classical algorithm
would do irreversibly).  However, the algorithm can be
performed on arbitrary quantum states (typically a uniform
superposition of input states in the Deutsch-Jozsa and
Bernstein-Vazirani algorithms).

A careful perusal of Equation~(\ref{bv-orig}) reveals two 
important facts about the Bernstein-Vazirani algorithm.
\begin{itemize}
\item[(a)] The
register qubit does not play any role in the algorithm. It is
used only  in the function evaluation step because the
oracle~(\ref{oracle1}) demands that we supply this extra qubit.
However, if we modify the oracle to 
\begin{equation}
\vert x \rangle_{\mbox{\tiny n-bit}}
\stackrel{U_{f}}{\longrightarrow}
(-1)^{f(x)} \vert x \rangle_{\mbox{\tiny n-bit}}
\label{new-oracle}
\end{equation}
we can implement everything on $n$ qubits.
Since the state of the last
qubit does not change, it can be considered redundant and we
can remove the one-qubit target register altogether. 

Although this oracle suffices to execute the Bernstein-Vazirani
algorithm, it  cannot give
us the value of $f(x)$ for a given $x$. Therefore, one can argue
that the connection with the original classical problem is lost
and one is solving an altogether different problem. In this
paper, we demonstrate that in the classical model based on
polarization of light beams, this problem can be
circumvented and we
can obtain the value of $f(x)$ from $x$ via a suitable
modification of the circuit.  We will come back to these subtle
points again in the next section.  
\item[(b)] 
It turns out that this version of the oracle is implementable
without requiring any entanglement for the case of the
Bernstein-Vazirani algorithm. 
The modified oracle~(\ref{new-oracle}) can be
implemented without introducing any entanglement because
the unitary
transformation $U_{a}$ can be decomposed as a direct product
of single qubit operations.
\begin{eqnarray} U_{a} &=&
U^{(1)}_{a} \otimes U^ {(2)}_{a}  \otimes
\cdots U^{(n)}_{a}\nonumber 
\\
&=&(\sigma_z^1)^{a_1} \otimes (\sigma_z^2)^{a_2}\cdots
(\sigma_z^n)^{a_n} 
\label{factor}
\end{eqnarray}
where $\sigma_z^j$ is the Pauli operator acting on the $j$ th
qubit.
On an $n$-qubit eigen state  $\vert x
\rangle = \vert x_1\rangle \vert x_2\rangle 
\cdots \vert x_n\rangle$ labeled by the binary string $x$ the
action reduces to
\begin{equation}
U_a\equiv  (-1)^{x_1.a_1}\, (-1)^{x_2.a_2}\, \cdots (-1)^{x_n.a_n}
\label{factor1}
\end{equation}
\end{itemize}
\begin{figure}[h]
\unitlength=.50mm
\thicklines
\begin{picture}(100,100)(10,10)
\begin{boldmath}
\def\box{
\multiput(0,0)(0,12){2}{\line(1,0){12}}
\multiput(0,0)(12,0){2}{\line(0,1){12}}
\put(2,3){$H$}
}
\put(20,81){\box}
\put(20,61){\box}
\put(20,31){\box}
\multiput(0,86)(32,0){2}{\line(1,0){20}}
\multiput(0,66)(32,0){2}{\line(1,0){20}}
\multiput(0,36)(32,0){2}{\line(1,0){20}}
\put(92,81){\box}
\put(92,61){\box}
\put(92,31){\box}
\multiput(72,86)(32,0){2}{\line(1,0){20}}
\multiput(72,66)(32,0){2}{\line(1,0){20}}
\multiput(72,36)(32,0){2}{\line(1,0){20}}
\multiput(52,30)(0,65){2}{\line(1,0){20}}
\multiput(52,30)(20,0){2}{\line(0,1){65}}
\put(56,60){$U_a$}
\put(-15,84){$\vert 0 \rangle$}
\put(-15,64){$\vert 0 \rangle$}
\put(-12,55){$.$}
\put(-12,50){$.$}
\put(-12,45){$.$}
\put(-15,34){$\vert 0 \rangle$}
\put(125,84){$\vert a_1 \rangle$}
\put(125,64){$\vert a_2 \rangle$}
\put(128,55){$.$}
\put(128,50){$.$}
\put(128,45){$.$}
\put(125,34){$\vert a_n \rangle$}
\end{boldmath}
\end{picture}
\caption{
\label{bv-circuit}
Pictorial representation of the Bernstein-Vazirani algorithm
using a modified oracle on $n\/$ un-entangled qubits.  Initially
the qubits are set to be all in the $\vert 0 \rangle$ state. Each
box containing $H$ represents a Hadamard transformation and the
box $U_a$ represents the oracle.  By a single call to the oracle
sandwiched between the hadamard gates we arrive at the final
state $\vert a\rangle$ which on measurement reveals the binary
vector $a$.}
\end{figure}
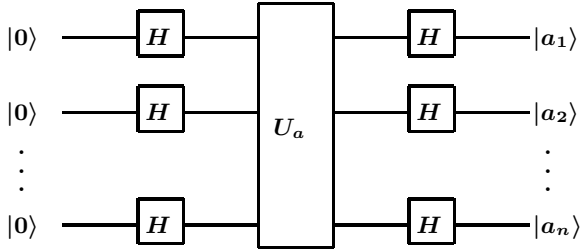
This simplified version of the 
Bernstein-Vazirani algorithm where only $n\/$ qubits
are used and we have separable states at all stages of
the implementation has been depicted in figure~(\ref{bv-circuit}).
All the implementations till date have been along the lines of
this circuit~\cite{djnmr,djoptics}.
\section{New optical implementation of the 
Bernstein-Vazirani algorithm} 
\label{new-algo}
It was shown in Section~(\ref{optics}) that 
$n\/$ classical beams of light
can be visualized as an $n\/$  qubit system and the action of
a non-entangling unitary transformation can be implemented via
a suitable combination of two quarter wave plates and a half wave
plate on each beam.  Although the set of unitaries that can be
implemented is limited, there is an added advantage that we can
clone these beams by using beam splitters.
The waves are classical and therefore there
is no problem in dividing the amplitude of a given polarization to
obtain two copies of the same polarization state. We 
will now use this property
of the model to implement the Bernstein-Vazirani 
algorithm in a new way and also
to make the modified oracle more powerful in terms of its
capacity to compute $f(x)$ from $x$.
\begin{figure}[h]
\unitlength=.50mm
\thicklines
\begin{picture}(100,100)(20,0)
\begin{boldmath}
\multiput(50,50)(0,40){2}{\line(1,0){20}}
\multiput(50,50)(20,0){2}{\line(0,1){40}}
\put(55,67){$U_a$}
\multiput(0,85)(70,0){2}{\line(1,0){50}}
\put(-16,83){`$\vert x_1 \rangle$'}
\put(-16,73){`$\vert x_2 \rangle$'}
\put(-8,63){.}
\put(-8,66){.}
\put(-8,69){.}
\put(-16,54){`$\vert x_n \!\rangle$'}
\multiput(0,75)(70,0){2}{\line(1,0){50}}
\multiput(0,55)(70,0){2}{\line(1,0){50}}
\multiput(120,85)(0,-10){2}{\line(1,0){10}}
\put(100,55){\line(1,0){30}}
\put(133.7,85){\line(1,0){13.3}}
\put(133.7,75){\line(1,0){11.3}}
\put(133.7,55){\line(1,0){15.3}}
\put(147,85.3){\line(0,-1){15.5}}
\put(145,75.3){\line(0,-1){7.5}}
\put(149,54.8){\line(0,1){6.4}}
\put(147,70){\line(1,0){14}}
\put(145,68){\line(1,0){17}}
\put(149,61){\line(1,0){13}}
\multiput(8,85)(102,0){2}{\line(0,-1){60}}
\multiput(15,75)(88,0){2}{\line(0,-1){40}}
\multiput(30,55)(59,0){2}{\line(0,-1){10}}
\put(8,25){\line(1,0){102}}
\put(15,35){\line(1,0){88}}
\put(30,45){\line(1,0){59}}
\multiput(130,88.6)(0,-10){2}{\line(0,-1){7.2}}
\put(130,58.6){\line(0,-1){7.2}}
\put(130,55){\oval(8,7)[r]}
\put(130,85){\oval(8,7)[r]}
\put(130,75){\oval(8,7)[r]}
\put(160,65){\oval(9,12)[r]}
\put(160,64){\oval(20,17)[tr]}
\put(160,66){\oval(20,17)[br]}
\put(158,65){\oval(9,12)[r]}
\put(155,75){XOR}
\put(155,50){$f(x)$}
\put(120,87){D$_1$}
\put(120,77){D$_2$}
\put(120,57){D$_n$}
\put(133,88){$\scriptstyle x_{\mbox{\tiny 1}}.a_{\mbox{\tiny 1}}$}
\put(133,78){$\scriptstyle x_{\mbox{\tiny 2}}.a_{\mbox{\tiny 2}}$}
\put(133,58){$\scriptstyle x_n.a_n$}
\thinlines
\multiput(13,80)(7,-10){2}{\line(-1,1){10}}
\put(8,87){BS$_{1}$}
\put(15,77){BS$_{2}$}
\put(95,87){BS$_{1}^{\prime}$}
\put(88,77){BS$_{2}^{\prime}$} 
\multiput(105,80)(-7,-10){2}{\line(1,1){10}}
\put(35,50){\line(-1,1){10}}
\put(28,58){BS$_{n}$}
\put(75,58){BS$_{n}^{\prime}$}
\put(84,50){\line(1,1){10}}
\multiput(10,20)(7,10){2}{
\line(-1,1){10}
\multiput(2,0.1)(-1,1){11}{\line(1,0){1}}
}
\multiput(105,20)(-7,10){2}{\line(1,1){10}
\multiput(-10,0.1)(1,1){11}{\line(1,0){1}}
}
\put(35,40){\line(-1,1){10}}
\multiput(35,40.1)(-1,1){11}{\line(-1,0){1}}
\put(84,40){\line(1,1){10}}
\multiput(84,40.1)(1,1){11}{\line(1,0){1}}
\end{boldmath}
\end{picture}
\label{classical-circuit}
\caption{Optical circuit to (a) implement the Bernstein-Vazirani algorithm in a new
way and (b) to compute $f(x)$ from $x$. BS's represent 50/50
beam splitters, the corner elements are mirrors and D's are
light detectors. The XOR gate is
implemented on pairs of bits till we are left with only
a one bit result.}
\end{figure}
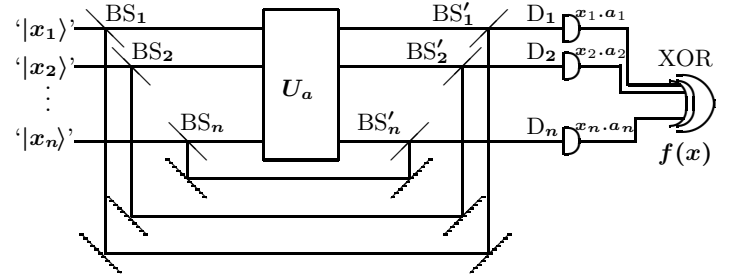
A notation similar to quantum mechanics 
is used in which single
quotations marks will be used around ket vectors for describing
the polarization states of light beams, where `$\vert x_j\rangle$'
represents $x$-polarization if $x_j=0$ and 
$y$-polarization if $x_j=1$. Each beam splitter in the circuit
splits the beam into two, keeping the polarization state of both the
beams identical to the original polarization. The intensity of
the split beams is half that of the original beam.

We now follow the circuit described in
Figure~(\ref{classical-circuit}) to arrive at our 
results.  Consider the input state labeled by the binary
vector $x$ with its bits given by  $x_{1}, x_{2} \cdots
x_{n}$. We represent it by a polarization state
`$\vert x_{1}\rangle$'`$\vert x_2\rangle$'$\cdots$ `$ \vert
x_n \rangle$' where each beam has an $x$ or $y$ polarization
depending upon the corresponding bit being $0$ or $1$.

Each beam goes through an identical set of operations.
Consider the $j$th beam.  The initial state of this beam is
`$\vert x_j\rangle$' and after the beam splitter we have two
copies of the same state (classical cloning of polarization
states).  The oracle acts on one of the copies and converts
it via the unitary transformation $U_a^j=(-1)^{x_j.a_j}$;
the other copy does not undergo any change. Both these
copies are brought together and mixed at the beam splitter
BS$_j^{\prime}$ and the intensity is measured at the
detector $D_j$.

\begin{eqnarray}
{\rm `}\vert x_j\rangle{\rm '}\longrightarrow\begin{array}{ccc}
\frac{1}{\sqrt{2}}{\rm
`}\vert x_j\rangle{\rm '}&&{\rm Transmitted}\\ \\
-\frac{1}{\sqrt{2}}{\rm `}\vert x_j\rangle{\rm '} &&{\rm reflected}
\end{array}
\end{eqnarray}

The transmitted component then undergoes the action of the oracle
unitary (Equations~(\ref{factor}) and (\ref{factor1})) which for 
the $j$th qubit acts via
$U_a^j=\sigma_z^{a_J}=(-1)^{x_j.a_j}$. The state of the
beam is 

\begin{equation}
\frac{1}{\sqrt{2}}{\rm `}\vert x_j\rangle{\rm '}\stackrel{U_a^j}{\longrightarrow}
\frac{1}{\sqrt{2}}(-1)^{x_j.a_j} {\rm `}\vert x_j\rangle{\rm '}
\end{equation}

As is clear from Equation~(\ref{biref}), the implementation
of $U_a^j$ on the $j$th beam is straightforward and is a
polarization dependent phase shift corresponding to a
single half wave plate with $\phi=0$ and $\eta=\pi$
~\cite{phase}. 

Finally, this beam meets the other beam (the one 
that did not
undergo the oracle unitary) at the beam splitter $BS_j$, where they
interfere to give the state of the beam moving towards
detector $D_j$ 

\begin{equation}
\frac{1}{2}(-1)^{x_j.a_j} {\rm `}\vert x_j\rangle{\rm '}-
\frac{1}{2}{\rm `}\vert x_j\rangle{\rm
'}=\frac{1}{2}((-1)^{x_j.a_j} -1) {\rm `}\vert x_j\rangle{\rm '} 
\label{phase}
\end{equation}

The negative sign in Equation~(\ref{phase}) implies that the
beam which does not pass through the oracle acquires an
extra phase factor of $\pi$, which can be easily arranged.
After interference, the amplitude and hence the intensity at
the detector $D_j$ is zero if $x_j.a_j$ is zero.  On the
other hand, if $x_j.a_j$ is one the intensity at the
detector is $\frac{1}{4}$ (assuming that we started with a
beam of unit intensity).{\em  This happens for all the beams and
therefore each detector measures the corresponding
$x_j.a_j$}.

\subsection{To find the $n$ bit string `$a$'}

We can find $x_j.a_j$  separately for all $j$'s with this
simple interferometric arrangement. The computation of the
binary string $a$ is now straightforward. If we choose a
special input state with $x_j=1$ for all the $j \in
\{1,2,\cdots n\}$ then the detectors measure the
corresponding $a_j$ and therefore we are able to compute the
string $a$. This is quite different from the quantum version
of the Bernstein-Vazirani algorithm where we use Hadamard
gates to create superpositions. As a matter of fact, the
scheme with Hadamard gates described in
Figure~(\ref{bv-circuit}) can also be implemented in our
model with polarization qubits.
\subsection{Computing $f(x)$ for a given $x$}
In order to compute $f(x)$ for a given $x$, the appropriate
polarization state representing the $n$-bit input $x$ is chosen.
The outputs from all the detectors are fed into  a pair-wise XOR
gate to compute addition modulo 2 (the XOR is applied to pairs of
inputs until we are left with only one output).  This process
amounts to computing $f(x)= x_1.a_1\oplus x_2.a_2 \cdots
\oplus x_n.a_n$.  We can thus compute $f(x)$ for any given $x$.
\section{Concluding Remarks}
\label{conclusion}
We have described a  classical optical scheme to implement
the Bernstein-Vazirani algorithm. This scheme is entirely
classical as we have used only 'classical qubits' (based on the
polarization states of light beams), and passive optical
elements such as 
detectors, beam splitters, phase shifters and mirrors. The number of
components needed to implement the algorithm increases
linearly with the number of input beams. We have explicitly cloned
the input and interfered it again with the part which undergoes
the oracle unitary, in order to 
solve the Bernstein-Vazirani problem. This scheme does not 
require the implementation of any Hadamard gates. We have also shown
through our interference arrangement that we can 
use the same oracle to compute
$f(x)$ for a given $x$. 

We believe that this analysis is a step in the direction
where information processors based on interference of waves are analyzed
in detail for their computation power.  These systems seem
to provide a model that is in-between the classical
computation model based on bits and a fully quantum
computer. The computational power is also likely to be
in-between the two models (these issues will be discussed
elsewhere).


\begin{thebibliography}{99}
\bibitem{phase}We have to introduce an overall phase to bring
the final output to the desired form. However, that can be done
trivially and is hence not shown explicitly.
\end{thebibliography}
\end{document}